# P-type behavior of CrN thin films by control of point defects

Arnaud le Febvrier[1]*, Davide Gambino[1]*, Fabien Giovannelli[2], Babak Bakhit[1], Simon Hurand[3], Gregory Abadias[3], Björn Alling[1], Per Eklund[1]

*Corresponding authors: arnaud.le.febvrier@liu.se, davide.gambino@liu.se

## Affiliation

[1] *Department of Physics, Chemistry and Biology (IFM), Linköping University, SE-58183 Linköping, Sweden*

[2] GREMAN UMR-CNRS 7347, Université de Tours, INSA Centre Val de Loire *15 rue de la chocolaterie, CS 2903, 41029 Blois Cedex, France*

[3] *Institut Pprime, Département de Physique et Mécanique des Matériaux, UPR 3346, CNRS-Université de Poitiers-ENSMA, 11 Boulevard Marie et Pierre Curie, - TSA 41123, 86073 Poitiers Cedex 9, France*

**Abstract**

We report the results of a combined experimental and theoretical study on nonstoichiometric $CrN_{1+\delta}$ thin films grown by reactive magnetron sputtering on c-plane sapphire and MgO (100) substrates in a $Ar/N_2$ gas mixture using different percentages of $N_2$. There is a transition from n-type to p-type behavior in the layers as a function of nitrogen concentration varying from 48 to 52 at. % in CrN films. The compositional change follows a similar trend for all substrates, with a N/Cr ratio increasing from approximately 0.7 to 1.06-1.11 by increasing percentage of $N_2$ in the gas flow ratio. As a result of the change in stoichiometry, the lattice parameter and the Seebeck coefficient increase together with the increase of N in $CrN_{1+\delta}$; in particular, the Seebeck value coefficient transitions from -50 $\mu V.K^{-1}$ for $CrN_{0.97}$ to +75 $\mu V.K^{-1}$ for $CrN_{1.1}$. Density functional theory calculations show that Cr vacancies can account for the change in Seebeck coefficient, since they push the Fermi level down in the valence band, whereas N interstitial defects in the form of $N_2$ dumbbells are needed to explain the increasing lattice parameter. Calculations including both types of defects, which have a strong tendency to bind together, reveal a slight increase in the lattice parameter and a simultaneous formation of holes in the valence band. To explain the experimental trends, we argue that both Cr vacancies and $N_2$ dumbbells, possibly in combined configurations, are present in the films. We demonstrate the possibility of controlling the semiconducting behavior of CrN with intrinsic defects from n- to p-type, opening possibilities to integrate this compound in energy-harvesting thermoelectric devices.

**Introduction**

Thermoelectric (TE) materials are promising candidates for energy-harvesting devices [1-3] that can convert heat into electricity when they are exposed to a temperature gradient. A thermoelectric device is composed of several legs electrically connected in series of n-type and p-type semiconductors [4,5]. The TE figure of merit ($zT$), a dimensionless quantity related to the TE efficiency, is defined as $zT = S^2 \sigma T/(k_e + k_p)$, where $S$, $\sigma$, $T$, $k_e$ and $k_p$ are the Seebeck coefficient, the electrical conductivity, the absolute temperature, the electronic thermal conductivity and lattice thermal conductivity (phonon), respectively. $zT$ can be optimized by maximizing the parameter $S^2$, called power factor, and minimizing the lattice thermal conductivity, which is the main contribution to the thermal conductivity in semiconductors. Among established thermoelectric materials, telluride-based materials typically show the highest $zT$ values [3]. However, for future production and/or use in commercial aspect, search for alternative materials is urgently needed because of the scarcity, toxicity, and low global production of tellurium [6-10]. To accommodate this demand, the search for new thermoelectric materials is largely focused on silicides, oxides, nitrides, and organic materials with elements which are more abundant and less toxic [7,11]. Among nitrides for thermoelectric applications, scandium- and chromium-based nitrides have been reported with promising thermoelectric properties [12-18].

Sc- and Cr-based nitride materials exhibit n-type conduction with promising properties for further integration in TE devices. However, for device integration both n-type and p-type materials are needed. P-type ScN can be fabricated by alloying/doping with Mg or Mn on Sc sites [10,13,19]. The power factor of *n*-type CrN is high, and this material has an intrinsically lower thermal conductivity than ScN [12,20,21]. The main mechanism that leads to lower thermal conductivity in CrN is related to its magnetic properties, since this compound undergoes a transition from an antiferromagnetic to a paramagnetic (PM) state at a critical temperature $T_c$ close to 280 K [22]: above $T_c$, scattering of phonons due to the magnetic disorder reduces even further the thermal conductivity as compared to other nonmagnetic rock-salt transition-metal nitrides [23].

First-principles modelling of materials in the PM state is not trivial. Approximation of the PM phase of a magnetic material with a nonmagnetic state leads to wrong results in some cases [24-27]. However, the PM phase can be modelled by density functional theory (DFT) calculations with the disordered local moment (DLM) approximation, in which the PM state is represented by disordered magnetic moments in the lattice [28-32]. Initially employed with the coherent potential approximation (CPA) [33], it has been then extended to supercell calculations [34], which allow to perform local lattice relaxations, a main requisite for the investigation of defects. In the past years, other methods to perform relaxations consistently in the PM state have been also developed [35,36].

Defects are important in transition-metal nitrides with the cubic rock-salt structure, susceptible to non-stoichiometry [15,37-40]. In the case of CrN, the presence of vacancies and/or other types of defects coupled with the narrow band gap of the intrinsic material varies the charge carrier density and thus provides a wide range of electrical properties from semiconductor to metallic-like conduction [41,42].

Several studies have reported that n-type $CrN_{1-\delta}$ exhibit relatively high $zT$ (0.1 - 0.35 in a temperature range from RT to 400 °C, with moderate electrical conductivity (1.7 – 350 mΩ.cm), low thermal conductivity (2 - 4 $W.m^{-1}.K^{-1}$), and high absolute values of the Seebeck coefficient (135 - 200 $\mu V.K^{-1}$) [15,43-45]. The n-type behavior of understoichiometric $CrN_{1+\delta}$ has been explained by DFT calculations to be induced by N vacancies, which shift the Fermi level towards the conduction band [38]. N interstitials in the form of $N_2$ dumbbells have been also shown to induce an n-type behavior in this compound, although the effect is weaker than observed with N vacancies. McGahay et al. showed that growing CrN thin films in the presence of oxygen leads to the formation of an oxynitride $Cr_{1-x/2}N_{1-x}O_x$ compound with random distribution of O atoms on the N sublattice accompanied by the formation of Cr vacancies [46]. By increasing O concentration, they observed an insulator-to-metal transition, followed by an additional transition back to an insulating state. For $x \leq 0.26$, the $Cr_{1-x/2}N_{1-x}O_x$ layers are n-type semiconductors with increasing number of carriers of this type. On the other hand, p-type behavior was observed for overstoichiometric CrN doped with Al ($Cr_{0.96}Al_{0.04}N_{1.17}$) film [47] and on films deposited by radio-frequency sputtering using relatively large portion of nitrogen as reactive gas, although no detailed film composition was reported in the latter case [48]. These different studies have shown the sensitivity of the electrical behavior of CrN with dopants or contaminants. To improve the thermoelectric properties of the potential p-type CrN, there is a need of fundamental understanding of the underlying origin of this p-type semiconductor behavior in CrN without any dopant. This may guide further studies on the control the electrical properties with dopants ultimately.

Therefore, with a motivation to develop both n-type and p-type CrN-based thermoelectric materials, ultimately required for TE devices, we demonstrate here the possibility to convert n-type CrN to p-type $CrN_{1+\delta}$ by controlling the stoichiometry, without the need for dopants. We show the effect of Cr vacancies combined with nitrogen interstitials on the thermoelectric properties of this $CrN_{1+\delta}$ compound. The different characteristics of the films synthesized by reactive magnetron sputtering were examined (crystal structure, morphology, elemental composition), and the thermoelectric properties evaluated from ambient temperature up to 525 °C. DFT calculations of PM $CrN_{1+\delta}$ are carried out with defects compatible with the N overstoichiometry on the Cr sublattice, on the N sublattice, and on both sublattices at the same time. These calculations indicate that Cr vacancies can induce the p-type behavior observed experimentally, although presence of $N_2$ dumbbells is needed to explain the expansion of the lattice parameter.

**Experimental and method**

Material synthesis:

330 nm $CrN_{1+\delta}$ thick films were deposited simultaneously onto c-plane sapphire, MgO (100) and $LaAlO_3$(100) (abbreviated LAO) substrates using a magnetically unbalanced magnetron sputtering system (base pressure ~ $10^{-5}$ Pa) equipped with confocal target configuration [49]. The Cr (99.95 % purity) target was 7.5-cm (3 inch) diameter water-cooled disk electrically connected to a DC power supply at a constant power of 200 W. The films were deposited using different $N_2$/Ar gas mixtures. The flow of nitrogen was kept constant at 20 sccm, while the flow of argon was changed from 6 to 30 sccm leading to different $N_2$ % of the total gas-flow from 77 % to 40 % and different working pressures from 0.25 to 0.40 Pa, respectively. The target was located at 18 cm from the substrate, which was mounted on a rotating sample holder coupled with a heating element. The nominal temperature was fixed at $T_s$ = 700 °C (corresponding approximately to 600°C at the substrate surface), and a substrate bias voltage of -30 V (DC equivalent) was applied during deposition using an RF power supply. The substrates were first cleaned using detergent steps, then 10 min using acetone in ultrasonic bath and repeated with ethanol and finally blown dry with a $N_2$-gun. The detergent steps are described elsewhere [50]. Prior deposition, an in-situ Ar etching was performed using a RF substrate bias of – 90 V for 2 minutes. The total deposition time was 3500 s.

Materials analysis:

The crystal structure of the films was determined by X-ray diffraction (XRD). The θ-2θ scans were recorded in an X'Pert PRO from PANalytical in a Bragg Brentano using a Cu $K_\alpha$ radiation with a nickel filter. Philips X'Pert-MRD with Cu K$_\alpha$ radiation was used determining with accuracy the cell parameter on MgO substrate using a hybrid mirror on the incidence beam path and a triple axis Ge 220 analyser on the diffracted beam path. The surface morphology of the films was examined using a scanning electron microscope (SEM, LEO Gemini 1550, Zeiss) operated at 5 KeV.

The elemental compositions of nitride films were determined by time-of-flight elastic recoil detection analysis (ToF-ERDA) in a tandem accelerator. ToF-ERDA was carried out with a 36 MeV $^{127}$I$^{8+}$ probe beam incident at 67.5° with respect to the sample surface normal, and recoils are detected at 45°. More details about the measurements are given in reference [51]. The elemental compositions were also determined using a wavelength dispersive spectrometer (WDS) unit from Oxford Instruments (High Wycombe, UK) attached to a JEOL 7001 TTLS scanning electron microscope (Tokyo, Japan) operated at 10 kV.

Several measurement setups were used to investigate the electrical and thermoelectric properties. The Seebeck coefficient was measured using two different setups. The first one at room temperature (RT) using a home-built thermoelectric measurement setup equipped with two Peltier heat sources for creating a temperature gradient in the sample, two K-type thermocouples for measuring the temperature. The two electrodes are made of Cu and are in contact with the sample in an area of 9 mm x 1 mm in which the K-type thermocouples are present and electrically isolated. The different voltages obtained for different ΔT (0 to 10 K) were measured with a Keithley 2001 multimeter and two thermometers. Room temperature resistivity measurements were performed in a Van der Pauw configuration using a Ecopia AMP-55 Hall measurement system. The in-plane Seebeck coefficient and the electrical resistivity were measured simultaneously under a low-pressure helium atmosphere (~ 0.09 MPa, purity 99.999 (H$_2$O < 3ppm and O$_2$ < 2ppm) using ULVAC-RIKO ZEM3 from RT up to 400 °C. The substrate contribution to the Seebeck coefficient and electrical resistivity is negligible, and the instrumental error is within 10 %.

Theoretical calculation:

Density functional theory (DFT) calculations were performed with the Vienna Ab initio Simulation Package (VASP) [52,53] with projector augmented-wave (PAW) [54,55] potentials and employing LDA exchange-correlation functional. Strong electron correlation was included with the LDA+U scheme as devised by Dudarev [56], employing a value U$^{eff}$ = 3 eV on the Cr 3d electrons which is known to describe well both the low and high temperature phases, together with reasonable electronic properties [34]. The energy cut-off for expansion of plane-waves was set to 400 eV, while sampling the first Brillouin zone with a 3×3×3 Monkhorst-Pack *k*-mesh. The criterion for convergence of electronic optimization was set to $10^{-3}$ eV/supercell. Spin-polarized calculations are carried out with collinear moments. To model the PM phase, we used the magnetic sampling method (MSM) [34], a supercell implementation of the DLM approach, which consists of employing many different magnetic configurations of randomly distributed up and down moments. The defective $CrN_{1+\delta}$ structures were relaxed in the DLM state with the DLM-relaxation method described in Ref. [35] using the defect free equilibrium volume.

The equilibrium lattice parameter of the structures with a single defect (a Cr vacancy or a $N_2$ dumbbell) was obtained from the equation of state, calculated with a supercell made of 2×2×2 repetitions of the conventional rock-salt cell (32 formula units) and employing five different volumes, using the atomic positions relaxed at the defect-free equilibrium volume. At each volume, we carried out MSM calculations with ten different magnetic configurations. The equation of state was calculated for each magnetic configuration separately and the equilibrium lattice parameter for the defective structures was then retrieved as an average of the equilibrium lattice parameters of each magnetic configuration. For the combined defects (Cr vacancy and $N_2$ dumbbell in the same supercell), we performed the same procedure but with a 3×3×3 supercell.

The calculation of formation and interaction energy of defects was carried out with a 3×3×3 supercell ($n = 108$ formula units) to minimize self-interaction of the defects. The energy of the defective supercells was obtained by MSM static calculations with 25 different magnetic configurations on the geometry resulting from DLM-relaxation of these 3x3x3 supercells. The total energy of the supercell was obtained as an average over all the magnetic configurations. The formation energy of the single or combined defect is defined as:

$$E^f(\text{Cr}_{n-v}\text{N}_{n+d}) = E(\text{Cr}_{n-v}\text{N}_{n+d}) - \left[\frac{n-v}{n}E(\text{Cr}_n\text{N}_n) + \frac{v+d}{2}E(\text{N}_2)\right], \qquad (1)$$

where $E(\text{Cr}_{n-v}\text{N}_{n+d})$ is the energy of the defective supercell, $E(\text{Cr}_n\text{N}_n)$ is the energy of the defect-free CrN 3×3×3 supercell, $n$ is the number of formula units in the defect-free supercell, $E(\text{N}_2)$ is the energy of a $N_2$ molecule, $v = \{0,1\}$ is an index that indicates the presence ($v = 1$) or absence ($v = 0$) of a vacancy in the Cr sublattice, and $d = \{0,1\}$ is a corresponding index for the N interstitial atom in the form of a $N_2$ dumbbell (notice the different sign for the $v$ and $d$ indices in the formula). We choose to employ the chemical potential of N for both types of defects because this is likely the relevant chemical potential in experiments [57]. The interaction energy between two defects is defined as:

$$E^i(\text{Cr}_{n-v}\text{N}_{n+d}) = E^f(\text{Cr}_{n-v}\text{N}_{n+d}) - E^f(\text{Cr}_{n-v}\text{N}_n) - E^f(\text{Cr}_n\text{N}_{n+d}), \qquad (2)$$

with negative (positive) energy indicating attractive (repulsive) interaction.

In addition, we calculated the density of state (DOS) of relevant supercells with defects in the PM phase as an average over all MSM magnetic configurations employed.

## Results and discussion

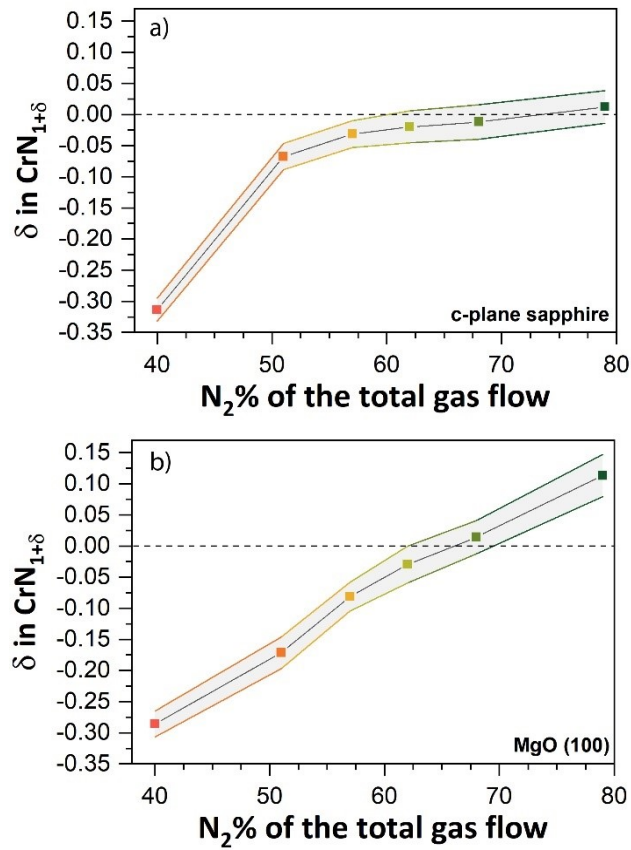

**Figure 1:** Film non-stoichiometry (from ToF-ERDA) represented as $\text{CrN}_{1+\delta}$ for the series of film deposited on c-plane sapphire (a) and MgO (100) (b) with different $N_2$ % of the total gas flow. The grey zone represents the error bar extracted from the ToF-ERDA analysis. Note here that the oxygen contamination is not taken into account in the calculation.

Figure 1 is a visual representation of the evolution of the film non-stoichiometry, denoted as $\text{CrN}_{1+\delta}$, measured by ToF-ERDA on c-plane sapphire and MgO (100) when the $N_2$ % of the total gas flow is varied. For all samples, the oxygen contamination in the films was estimated to be equal or less than 0.9 % with a lower oxygen content on c-plane sapphire with an average of 0.4 % as compared to 0.7 % on MgO (table S1 in the supplementary information). Differences in O contamination between the two series most probably originates from the different substrate, suggesting that diffusion from the substrate to the film may be more pronounced on MgO substrate. On all substrates and at low $N_2$ %, the films have the N/Cr ratio $\cong 0.7$, which increases to values higher than 1.0 (between 1.01 and 1.11 depending on the substrate) when the $N_2$ % increases. Although the films were deposited on the different substrates in the same batch, differences in elemental composition are observed between substrates. For example, the films on MgO (100) have a higher N/Cr ratio in the N-rich side of the series and a larger decrease in the N/Cr ratio is found when decreasing $N_2$ % of the total gas flow in comparison with the films deposited on c-plane sapphire. Within the error bar of the ToF-ERDA data analysis, the films deposited on c-plane sapphire at a $N_2$ % between 57 and 77 % can be considered stochiometric ($CrN_1$). Nevertheless, small differences are present with the nominal value varying from $CrN_{0.97}$ to $CrN_{1.01}$ when the films are deposited at 57 or 77 $N_2$ %. The films deposited using 51 $N_2$ % or less are understoichiometric with a minimum value of $\delta$ in $CrN_{1+\delta}$ at -0.32 (or $CrN_{0.68}$). On MgO (100), the films deposited between 62 and 68 $N_2$ % are stochiometric ($CrN_1$). The films deposited using 57 $N_2$ % or less are understoichiometric with a minimum value of $\delta$ in $CrN_{1+\delta}$ at -0.29 (or $CrN_{0.71}$). Note here that the measurements were performed using two different techniques which their results show good agreement. The details of the elemental composition analysis can be found in the supplementary information (table S1 and table S2). The oxygen content detected (1 % or less) was not included in the calculation of the ratio anion/cation in the NaCl structure of CrN (Fig. 1). If the oxygen was included in the calculation, all ratios anion/Cr would increase by an average of + 0.02 in the N-rich side of the series: for example, at the highest $N_2$ % and on MgO (100), the (N+O)/Cr becomes 1.13 instead of 1.11 without taking into account the oxygen.

The morphology of the films in the N-rich $CrN_{1+\delta}$ is similar within a given substrate series, with rounded triangular grains which tend to merge, forming elongated features on c-plane sapphire while a well-arranged pavement of square shaped grains is observed on MgO (100) when the $N_2$ % is higher than -0.03 $\leq \delta \leq$ + 0.11. The changes within the series on c-plane sapphire are relatively small compared to the series on MgO (100) and explained by the relatively small changes in the composition in the films. An overstoichiometric film ($\delta \geq$ -0.01) seems to consist of smaller grains than an understoichiometric film ($\delta \leq -0.03$). More details on the morphology are presented in the supplementary information (Fig. S3).

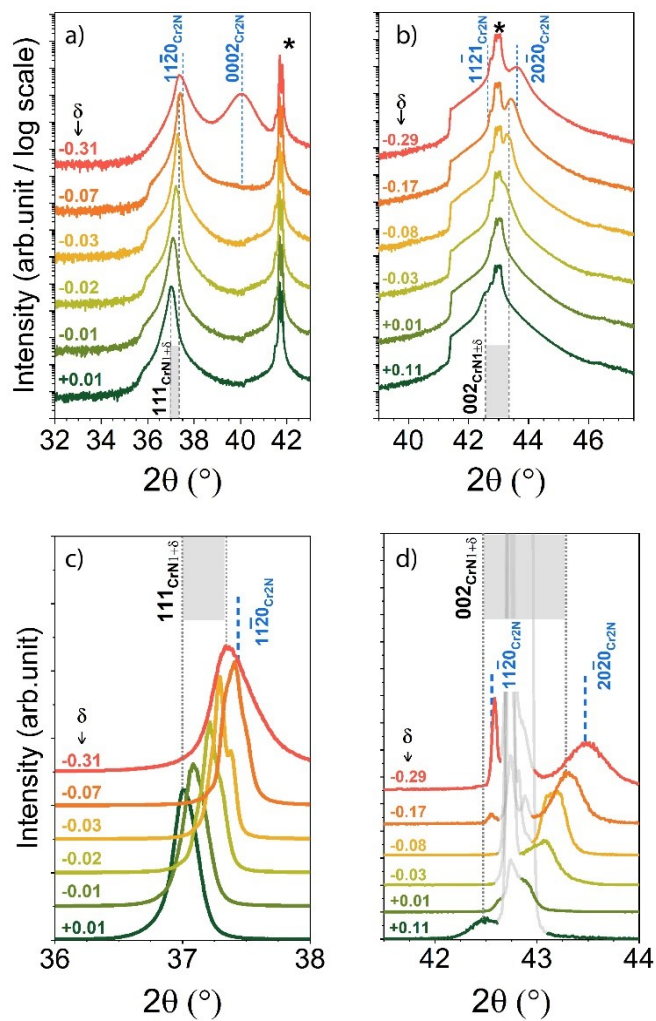

**Figure 2:** XRD patterns of CrN$_{1+\delta}$ films with different stoichiometries on a) and c) c-plane Al$_2$O$_3$, b) and d) MgO (100). Note that a) and b) are plotted using a log scale while c) and d) are with a linear scale. a), b) and c) are measurement performed in a Bragg Brentano configuration using a Cu K$_\alpha$ radiation with a nickel filter. d) is a HRXRD measurement performed using a Cu K$_{\alpha 1}$ radiation with an $\omega_{offset}$ = 0.1 – 0.2 °. Peaks marked with ∗ correspond to the substrate peaks: 0006 (Al$_2$O$_3$), 002 (MgO).

Figure 2 shows the XRD patterns of the CrN$_{1+\delta}$ films with different stoichiometries on different substrates (c-plane sapphire, MgO (100)) represented in log scale (a and b) and a magnified range in linear scale (c and d). The full XRD patterns (2θ: 10-90°) are presented in the supplementary information (Fig. S1). For the clarity of the figure 2d the substrate peak has been dimmed to visualize better the film peaks. The XRD pattern are also represented in supplementary information individually. Regardless of the substrate, the diffractogram of the CrN$_{1+\delta}$ films with $-0.03 \leq \delta \leq +0.11$ exhibits one diffraction peak belonging to the film depending on the N$_2$ % in the gas flow. This peak is identified as the one from the rock-salt CrN phase. Note here that the diffraction peak belonging to the film differs between 111 and 002, revealing a highly oriented film along the [111] or [001] directions of CrN on c-plane sapphire or on MgO (100), respectively. The orientation observed on c-plane sapphire is typical of an epitaxial growth which consists of twin domains in the cubic material thin film on c-plane sapphire [58]. On MgO (100) substrate, CrN grows following the so-called cube-on-cube epitaxial growth [59,60]. The term "epitaxial" is here used to describe the polycrystalline film composed of grains epitaxially oriented to the substrate crystal structure. Note here that the orientation of the films is consistent with the morphology of the N-rich CrN$_{1+\delta}$ films with triangular shape grains for a (111) oriented films and square shape grains for (001) oriented films. The orientation of the films, hence the surface energy, may influence the adsorption/reaction of nitrogen molecules on top the surface of the film during deposition. This influence seems to be more pronounced for (001) oriented films than for (111) oriented ones, as Figure 1 shows.

With CrN$_{1+\delta}$ films having $\delta \leq -0.07$, two other diffraction peaks are visible on c-plane sapphire at 2θ = 37.4° and 40.6° identified as the $1\bar{1}20$ and 0002 peak from Cr$_2$N phase, see Fig. 2(a). These peaks are clearly visible on the XRD pattern when $\delta = -0.31$, but can be also distinguished from an increase in the background intensity and a different shapes of the peak on the XRD pattern for the CrN$_{1+\delta}$ films with $\delta \leq -0.07$. For the series deposited on MgO, this secondary phase identified as Cr$_2$N is not clearly visible on a Bragg Brentano configuration measurement measured due to the high intensity of the substrate peak and the proximity to the film peak, Fig. 2(b). A θ-2θ measurement using a monochromatic Cu K$_{\alpha 1}$ and an ω offset of 0.1-0.2 ° allows to diminish the intensity of the 002 peak from MgO and reveal the film peak (Fig. S2 in supplementary information). Similarly to the films deposited on c-plane sapphire, the phase composition in the films varies from pure CrN$_{1+\delta}$ at high N$_2$ % to a mixture of CrN$_{1+\delta}$ and Cr$_2$N when $\delta \leq -0.08$. At the lowest amount of N$_2$ % in the gas flow ratio, the film is mainly composed of Cr$_2$N phase. On all substrates and upon increasing of N content in CrN$_{1+\delta}$, the CrN$_{1+\delta}$ peak position is shifted towards lower 2θ values. Therefore, increasing nitrogen in CrN$_{1+\delta}$ leads to an increase of the cell parameter.

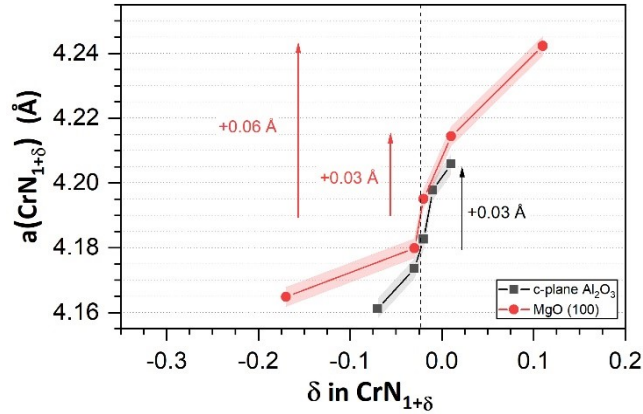

**Figure 3:** Cell parameter of cubic CrN$_{1+\delta}$ phase for films deposited on c-plane Al$_2$O$_3$, MgO (100). Values are extracted from the XRD peak corresponding to the main peak of CrN (001 or 111) depending on the substrate (see Fig. 2). The vertical line corresponds to the transition from negative to positive Seebeck coefficient (see Fig. 4). The vertical arrows represent the change of cell parameter on the same substrate compared to a film CrN$_{1+\delta}$ at which the transition from negative to positive Seebeck coefficient occurs.

Figure 3 displays the evolution of the cell parameter of the cubic CrN$_{1+\delta}$ phase for the different films as a function of $\delta$. Regardless of the substrate used, the CrN$_{1+\delta}$ phase has an increase of its cell parameter when the nitrogen content in the film increases. In this study, two references can be appropriate: i) The cell parameter of a stoichiometric film (CrN$_{1.00}$) around 4.20 ($\pm$ 0.005) Å which is comparable to the values reported in literature between 4.15 to 4.18 Å for CrN [43,45,60]; ii) the cell parameter of CrN$_{0.975}$ film at which the Seebeck coefficient switches from positive to negative values. The variation of the cell parameter is dependent of the nitrogen content in the film. A variation of $\delta$ = + 0.035 in CrN$_{1+\delta}$ (CrN$_{0.975}$ to CrN$_{1.01}$) lead to an increase of the cell parameter of +0.03 Å (+ 0.5%). An increase of 1.2 % of the cell parameter is observed for an increase of $\delta$ to + 0.11 in CrN$_{1+\delta}$.

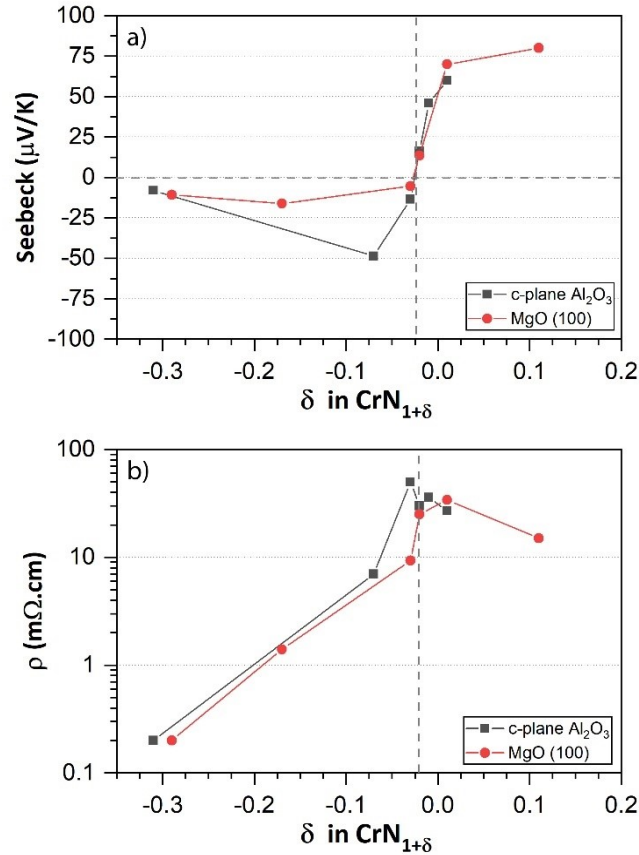

**Figure 4:** a) Seebeck coefficient RT and b) electrical resistivity values of the films deposited under different N$_2$ % of the total gas flow on different substrates (c-plane Al$_2$O$_3$, MgO (100)). The electrical properties are represented versus the nitrogen composition in CrN$_{1+\delta}$. The vertical line corresponds to the transition from negative to positive Seebeck coefficient.

Figure 4 displays the RT values of the Seebeck coefficient and the resistivity versus the nitrogen composition as $\delta$ in CrN$_{1+\delta}$. Regardless of the substrate, a trend is noticeable for the electrical properties of the film depending on the nitrogen content. The measured Seebeck coefficient at RT is positive (+ 50 to +80 µV.K$^{-1}$) when the films are overstoichiometric and becomes negative (- 25 to - 50 µV.K$^{-1}$) when the films are understoichiometric. The transition from positive to negative Seebeck coefficient occurs at $\delta \approx -0.025$ and not $\delta = 0$, which can be due to the presence of oxygen contamination (< 1%) and/or experimental uncertainties [46]. The electrical resistivity of the films stays relatively high, between 10 to 80 mΩ.cm, for the overstoichiometric films and reduces by one or two orders of magnitude for the understoichiometric ones. The films with $\delta$ = -0.30 have a low Seebeck coefficient (12 µV.K$^{-1}$) and low electrical resistivity (200 µΩ.cm), consistent with the physical characteristics of Cr$_2$N [61].

Figure 5 represents the thermoelectric properties measurements (Seebeck coefficient and electrical resistivity) from RT to 400°C for the films deposited on c-plane sapphire and on MgO (100).

The trend of the Seebeck coefficient and the electrical resistivity within a substrate series is similar as the one observed on Figure 4 where the measurement was performed on a different setup and only at RT. With the exception of the film with δ = -0.30, the electrical resistivity of the films decreases when the temperature increases revealing semiconductor behavior. The $CrN_{1+\delta}$ films with δ = -0.30 (corresponding to $CrN_{0.7}$) have their electrical resistivities increasing when the temperature increases; hence, a metallic behavior with a relatively low resistivity (0.2 mΩ.cm) consistent with the properties of $Cr_2N$ [61]. On both substrates, the Seebeck coefficient values tend to increase up to a maximum attained around 200-250 °C and then decreases slowly. Note that the film with δ of +0.11 on MgO (100) exhibits higher Seebeck coefficient values, and a maximum is observed at higher temperature (350 °C) compared to the other films.

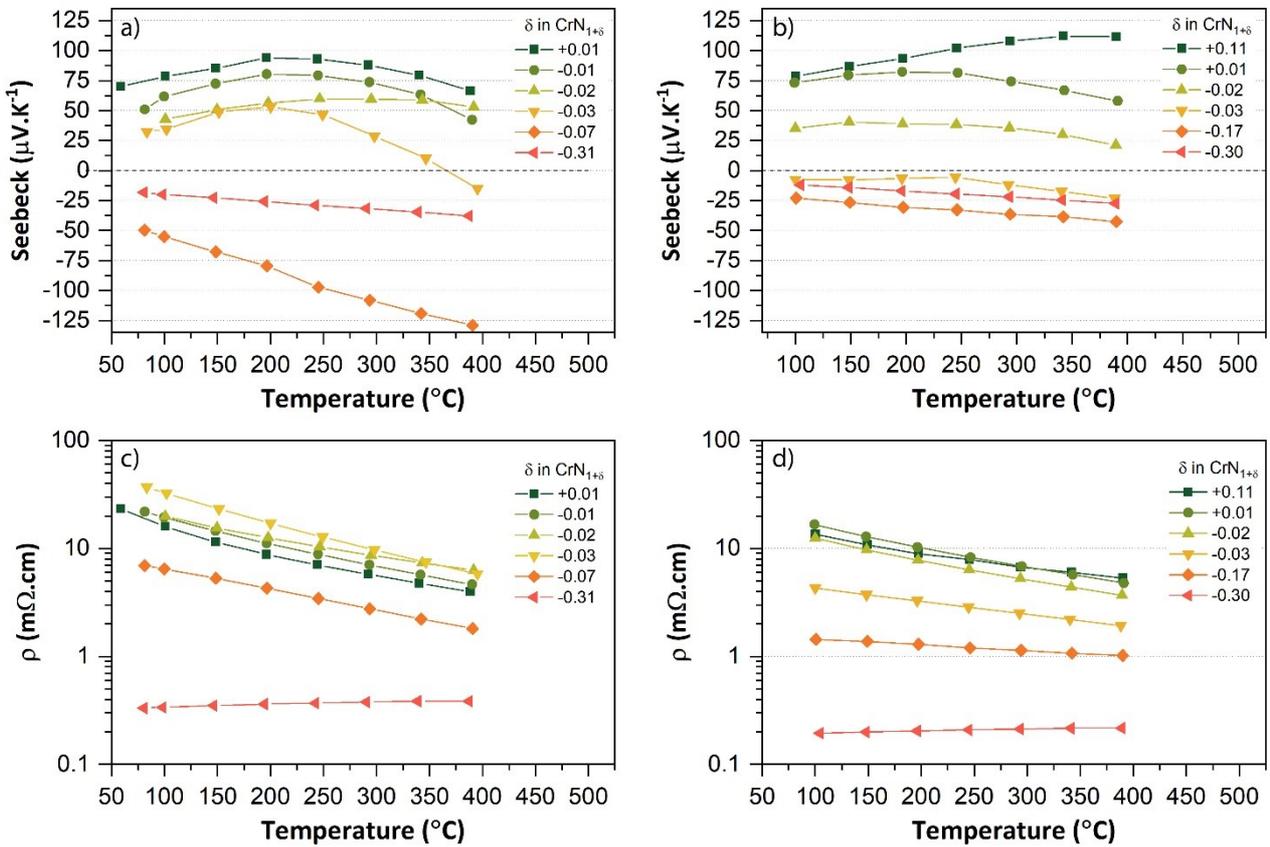

**Figure 5:** Temperature-dependence of the Seebeck coefficient (a, b) and electrical resistivity (c, d) of the films with different compositions: a) and c) on c-plane $Al_2O_3$, b) and d) on MgO (100)

The Seebeck coefficient of semiconductors can be described by the Mott-Boltzmann formalism from a relatively simple model with energy-independent scattering approximation and parabolic band [4]. The Seebeck coefficients ($S_n$) for the majority and minority charge carriers are defined as followed:

$$S_n = \frac{8\pi^2 k_B^2 T}{3eh^2} m^* \left(\frac{\pi}{3n}\right)^{2/3} \quad (3)$$

where n corresponds to the carrier concentration of electron (or p for holes), $m^*$ is the effective mass, T the temperature and h is Planck's constant [4,5]. For intrinsic semiconductors, both conduction and valence bands may play a role in the transport. Thus, the total Seebeck ($S_T$) coefficient is given by [5]:

$$S_T = \frac{S_n \sigma_n + S_p \sigma_p}{\sigma_n + \sigma_p} \quad (4)$$

The charge carriers (electrons (n) and holes (p)) would then compete and cancel out the induced Seebeck voltage. This is in analogy with metals, which have low Seebeck coefficients, and an extreme case is $Ti_3SiC_2$ with S = 0 over a wide temperature range [62-64]. In the present case, both p-type and n-type behavior were observed depending on the composition with a transition at room temperature happening at δ = -0.03 in $CrN_{1+δ}$. The temperature evolution of the Seebeck coefficient values for samples having low absolute Seebeck values (δ ≈ -0.03) can come from more complex phenomena with variation of both types of charge carriers (electrons and holes) with respect to both concentrations and mobilities when the temperature increases.

A theoretical approach was used to understand the underlying cause of this transition in the Seebeck values from negative to positive or from n-type to p-type semiconductor behavior as a function of the nitrogen content. DFT calculations were used to compute the formation energy of some possible defects responsible for nitrogen overstoichiometry in cubic $CrN_{1+δ}$ phase. The defects considered are the Cr vacancy, the N interstitial, and the N antisite. The N interstitial defect in CrN has been previously [38] shown to be most stable in a dumbbell configuration, i.e., a $N_2$ dimer placed with its center on a lattice site and aligned along a <111> direction. The formation energies of these defects are reported in Table 1. Cr vacancy is found to have the lowest formation energy of 2.48 eV. Formation energies in general define the relative stability of different defects in equilibrium conditions; however, the deposition technique employed in this study is far from equilibrium; therefore, thermodynamic reasoning alone cannot establish the prevalence of one defect over the other in the as-deposited films. For this reason, we cannot rule out the presence of both Cr vacancies and N dumbbells. Calculations with a combination of Cr vacancy and $N_2$ dumbbells were also taken into account, and we observe that the formation energy of these complexes is generally lower than the N antisite (see Table 1), which we therefore neglect in the rest of the discussion. We consider four configurations of Cr vacancy + N₂ dumbbell defects, which are shown in Fig. 6: in the first two cases [a) and b) in Fig. 6], the dumbbell is positioned on a lattice site nearest neighbor to the Cr vacancy, and aligned either along [110] in Fig. 6a) or [100] in Fig. 6b). In the third and fourth configurations (Fig 6c and 6d), the N₂ dumbbell is in third nearest neighbor position to the Cr vacancy with different orientations, where in the former case the dumbbell is oriented toward the vacancy. As Table 1 shows, the formation energy depends on the defect configuration, with energy of formation varying from 4.75 to 6.07 eV. The combined defect with highest formation energy displays a weak interaction energy as compared to the other defects, as can also be realized from Fig. 6d), since in this case the dumbbell is not directed towards the Cr vacancy. Finally, we see that the interaction energies of the defects are all negative, indicating attraction between the Cr vacancy and N₂ dumbbell; therefore, during the deposition process, we can argue that if these defects are formed near each other, they will attract and form a more stable configuration.

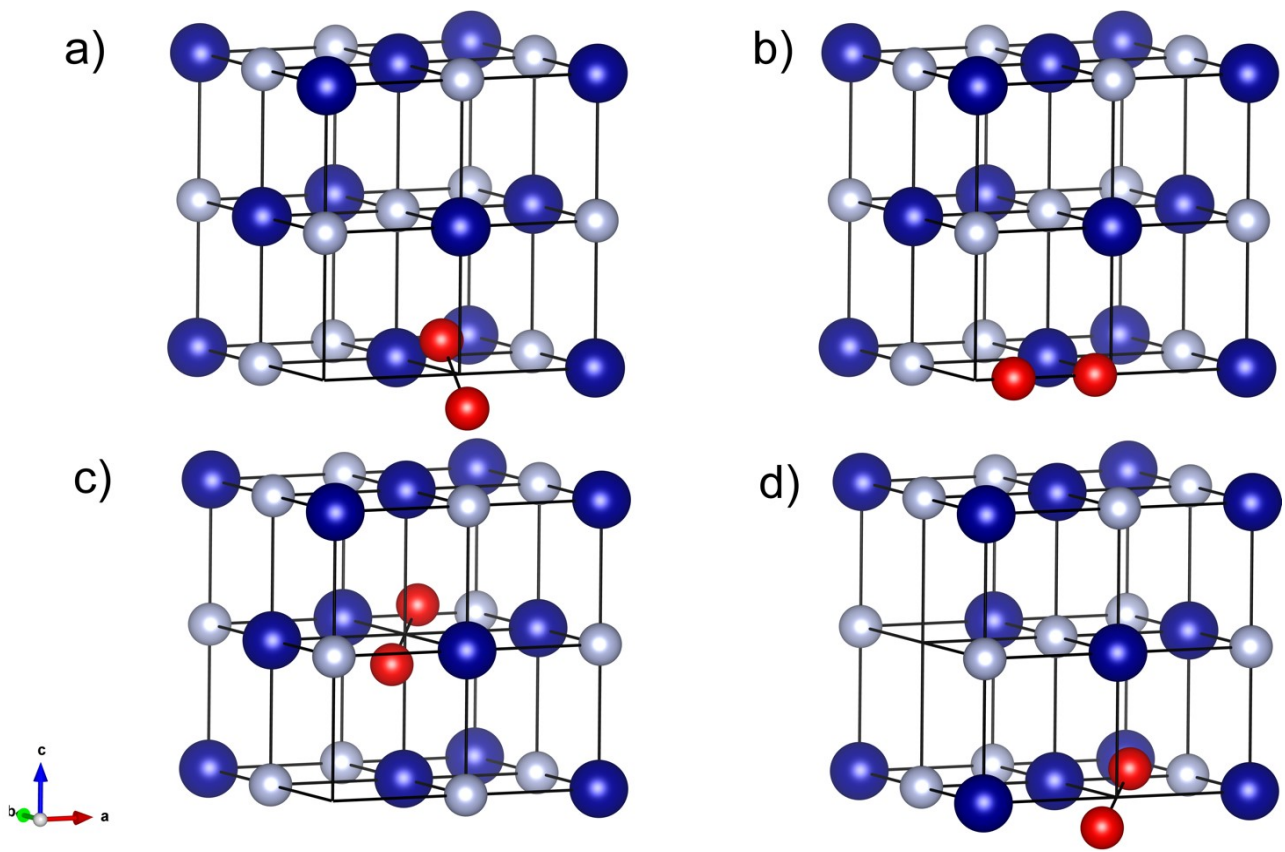

**Figure 6:** Structural models of the combined defects Cr vacancy-N₂ dumbbell in a single conventional rock-salt cell. Cr atoms are in blue, N atoms in grey and the N atoms in the dumbbell in red. Notice the missing Cr atom in position (0,0,0) in all models except in d), where the vacancy is at (0,1/2,1/2). The atoms on ideal lattice positions are shown without relaxation, whereas the N₂ dumbbell is shown as relaxed. From a) to d), the N₂ dumbbell is oriented in directions [110], [100], [111] and [111], respectively.

**Table 1:** Formation energy of the Cr vacancy ($Cr^v$), $N_2$ dumbbell ($N_2^d$), and N antisite ($N^a$) defects, and interaction energies of combined Cr vacancy and $N_2$ dumbbell ($Cr^v$+ $N_2^d$)) defects in the different configurations shown in Fig. 7. For the combined defects, the notation (a), (b), (c) and (d) refers to Fig. 6.

| Type of defects | Formation energy (eV) | Interaction energy (eV) |
|---|---|---|
| $Cr^v$ | 2.48 | - |
| $N_2^d$ | 3.61 | - |
| $N^a$ | 6.91 | - |
| $Cr^v$+ $N_2^d$ (a) | 4.75 | - 1.34 |
| $Cr^v$+ $N_2^d$ (b) | 5.23 | - 0.86 |
| $Cr^v$+ $N_2^d$ (c) | 5.88 | - 0.21 |
| $Cr^v$+ $N_2^d$ (d) | 6.07 | - 0.02 |

**Table 2:** Equilibrium lattice parameter for defect-free CrN, CrN with one Cr vacancy ($CrN_{1.03}$), and with one $N_2$ dumbbell ($CrN_{1.03}$). Note here that a supercell made of 2×2×2 or 3×3×3 repetitions of the conventional rock-salt cell is used for the single defect or the combined defect calculation, respectively.

| Type of defects | Labeled | N/Cr | Lattice parameter (Å) | Relative change (%) |
|---|---|---|---|---|
| Defect free | CrN | 1.00 | 4.129 | 0 |
| $Cr^v$ | $Cr_{1-x}N$ | 1.03 | 4.118 | - 0.3 |
| $N_2^d$ | $CrN_{1+y}$ | 1.03 | 4.159 | + 0.7 |
| $Cr^v$+ $N_2^d$ (a) | $Cr_{1-x}N_{1+y}$ | 1.02 | 4.136 | + 0.2 |

Table 2 reports the computed equilibrium lattice parameters for defective (Cr vacancy and the $N_2$ dumbbell and the most stable configuration of the combined defect) and defect-free CrN cubic structures. Comparatively to the defect-free structure, it is found that the $N_2$ dumbbell leads to an expansion of 0.7% of the lattice parameter, whereas the Cr vacancy induces a reduction by 0.3%; however, the combined defect seems to lead to a small increase of the lattice parameter as well, consistent with the experimental results with a variation of + 0.5 to + 1.2 %.

From previous studies [38], it is known that N vacancies in CrN can induce n-type behavior in the DOS. The same behavior is argued for in the case of the $N_2$ dumbbell, although with a much weaker strength. Figure 7 displays the DOS for the single defects and for the most stable configuration with the two defects combined [figure 6a]. We see that in the present cases with the Cr vacancy, the valence band is shifted to higher energy, leading to a p-type behavior, as observed in the present experiments. The $N_2$ dumbbell alone affects the DOS very weakly, shifting the valence band to lower energies as compared to the defect free system, whereas the Cr vacancy shows a more pronounced effect on the shift of the same band. The combination of the two defects is somewhat in between, but with character more similar to the Cr vacancy. These results indicate that the $N_2$ dumbbells alone, which can account for expansion of the lattice parameter as a function of $N_2$ content in the plasma, cannot be responsible for the p-type behavior seen in the experiments, but a combination of $N_2$ dumbbells and Cr vacancies could instead account for the positive Seebeck coefficient together with an expanded lattice parameter. It is also possible that upon higher concentration of defects more complex defects could lead to even larger effects on the electronic structure. Nonetheless, the results for highest non-stoichiometry are still compatible with the investigated defects and a detailed explanation of the deviation from linear behavior in the Seebeck coefficient and lattice parameter as a function of δ is beyond the scope of this work.

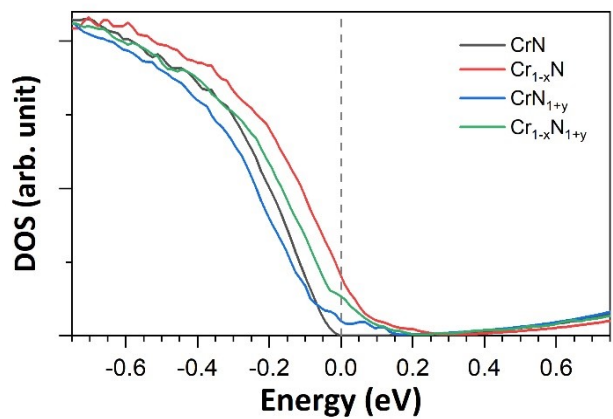

**Figure 7:** Electronic DOS around the Fermi level calculated with DFT on defect-free CrN, CrN with a Cr vacancy ($Cr_{1-x}N$), CrN with a $N_2$ dumbbell ($CrN_{1+y}$), and with the most stable configuration of the combined defects ($Cr_{1-x}N_{1+y}$), shown in Fig. 6a). The zero of the energy is the Fermi level.

## Conclusion

In conclusion, we show the possibility of synthesizing nonstoichiometric $CrN_{1+\delta}$ thin films ranging from n- to p-type behavior by increasing the N content. We explain this behavior with first principles calculations of intrinsic defects, which show that Cr vacancies induce the p-type behavior and $N_2$ dumbbells increase the lattice parameter, as observed in experiments. These two defects also show a strong interaction energy, suggesting that they attract each other and form complexes of defects in the system. Temperature-dependent measurements show also that the overstoichiometric $CrN_{1.11}$ films retain their large positive value of the Seebeck coefficient up to 400°C. This study is important for the development of thermoelectric devices based on CrN without the need of doping with foreign elements.

## Acknowledgements

The authors acknowledge support from the Knut and Alice Wallenberg Foundation through the Wallenberg Academy Fellows program (grant no. KAW 2020.0196), the Swedish Research Council (VR) under project grants 2016-03365 and 2021-03826, the Swedish Government Strategic Research Area in Materials Science on Functional Materials at Linköping University (Faculty Grant SFO-Mat-LiU No. 2009 00971), and the Swedish Energy Agency under project 46519-1. The authors also acknowledge support from the Swedish research council VR-RFI (2019-00191) for the Accelerator based ion-technological center and from the Swedish Foundation for Strategic Research (contract RIF14-0053) for the tandem accelerator laboratory in Uppsala University. BA acknowledge funding from the Swedish Foundation for Strategic Research through the Future Research Leaders 6 program, Grant No. FFL 15-0290, from the Swedish Research Council (VR) through Grant No. 2019-05403, and from the Knut and Alice Wallenberg Foundation (Wallenberg Scholar Grant No. KAW-2018.0194). The computations were enabled by resources provided by the Swedish National Infrastructure for Computing (SNIC) at NSC partially funded by the Swedish Research Council through Grant Agreement No. 2018-05973

## References

[1] F. J. DiSalvo, Science **285**, 703 (1999).
[2] J. P. Heremans, M. S. Dresselhaus, L. E. Bell, and D. T. Morelli, Nat Nano **8**, 471 (2013).
[3] X. Zhang and L.-D. Zhao, J. Materiomics **1**, 92 (2015).
[4] G. J. Snyder and E. S. Toberer, Nat Mater **7**, 105 (2008).
[5] H. J. Goldsmid, *Introduction to Thermoelectricity* (Springer, Veralg-Berlin Heidelberg, 2010), Mater. Sci.

[6] M. W. Gaultois, T. D. Sparks, C. K. H. Borg, R. Seshadri, W. D. Bonificio, and D. R. Clarke, Chem. Mater. **25**, 2911 (2013).
[7] J. He, Y. Liu, and R. Funahashi, J. Mater. Res. **26**, 1762 (2011).
[8] P. V. Burmistrova, J. Maassen, T. Favaloro, B. Saha, S. Salamat, Y. Rui Koh, M. S. Lundstrom, A. Shakouri, and T. D. Sands, J. Appl. Phys. **113**, 153704 (2013).
[9] M. Zebarjadi, Z. Bian, R. Singh, A. Shakouri, R. Wortman, V. Rawat, and T. Sands, J. Electron. Mater. **38**, 960 (2009).
[10] B. Saha, M. Garbrecht, J. A. Perez-Taborda, M. H. Fawey, Y. R. Koh, A. Shakouri, M. Martin-Gonzalez, L. Hultman, and T. D. Sands, Appl. Phys. Lett. **110**, 252104 (2017).
[11] G. Kieslich, G. Cerretti, I. Veremchuk, R. P. Hermann, M. Panthöfer, J. Grin, and W. Tremel, Phys. Status Solidi A **213**, 808 (2016).
[12] P. Eklund, S. Kerdsongpanya, and B. Alling, J. Mater. Chem. C **4**, 3905 (2016).
[13] S. Kerdsongpanya, B. Alling, and P. Eklund, Phys. Rev. B **86**, 195140 (2012).
[14] S. Kerdsongpanya, N. V. Nong, N. Pryds, A. Žukauskaitė, J. Jensen, J. Birch, J. Lu, L. Hultman, G. Wingqvist, and P. Eklund, Appl. Phys. Lett. **99**, 232113 (2011).
[15] C. X. Quintela, J. P. Podkaminer, M. N. Luckyanova, T. R. Paudel, E. L. Thies, D. A. Hillsberry, D. A. Tenne, E. Y. Tsymbal, G. Chen, C.-B. Eom, and F. Rivadulla, Adv. Mater. **27**, 3032 (2015).
[16] C. X. Quintela, F. Rivadulla, and J. Rivas, Appl. Phys. Lett. **94**, 152103 (2009).
[17] A. S. Botana, V. Pardo, and W. E. Pickett, Phys. Rev. Appl. **7**, 024002 (2017).
[18] N. Tureson, N. V. Nong, D. Fournier, N. Singh, S. Acharya, S. Schmidt, L. Belliard, A. Soni, A. l. Febvrier, and P. Eklund, J. Appl. Phys. **122**, 025116 (2017).
[19] H. A. AL-Brithen, H. Yang, and A. R. Smith, J. Appl. Phys. **96**, 3787 (2004).
[20] M. A. Gharavi, D. Gambino, A. le Febvrier, F. Eriksson, R. Armiento, B. Alling, and P. Eklund, Materials Today Communications **28**, 102493 (2021).
[21] B. Biswas, S. Chakraborty, O. Chowdhury, D. Rao, A. I. K. Pillai, V. Bhatia, M. Garbrecht, J. P. Feser, and B. Saha, Phys. Rev. Materials (2021).
[22] L. M. Corliss, N. Elliott, and J. M. Hastings, Phys. Rev. **117**, 929 (1960).
[23] I. Stockem, A. Bergman, A. Glensk, T. Hickel, F. Körmann, B. Grabowski, J. Neugebauer, and B. Alling, Phys. Rev. Lett. **121**, 125902 (2018).
[24] F. Rivadulla, M. Bañobre-López, C. X. Quintela, A. Piñeiro, V. Pardo, D. Baldomir, M. A. López-Quintela, J. Rivas, C. A. Ramos, H. Salva, J.-S. Zhou, and J. B. Goodenough, Nature Materials **8**, 947 (2009).
[25] B. Alling, T. Marten, and I. A. Abrikosov, Nature Materials **9**, 283 (2010).
[26] J. Varignon, M. Bibes, and A. Zunger, Nat. Commun. **10**, 1658 (2019).
[27] F. Körmann, A. Dick, B. Grabowski, T. Hickel, and J. Neugebauer, Phys. Rev. B **85**, 125104 (2012).
[28] J. Hubbard, Phys. Rev. B **19**, 2626 (1979).
[29] J. Hubbard, Phys. Rev. B **20**, 4584 (1979).
[30] J. Hubbard, Phys. Rev. B **23**, 5974 (1981).
[31] H. Hasegawa, J. Phys. Soc. Jpn. **46**, 1504 (1979).
[32] H. Hasegawa, J. Phys. Soc. Jpn. **49**, 178 (1980).
[33] B. L. Gyorffy, A. J. Pindor, J. Staunton, G. M. Stocks, and H. Winter, Journal of Physics F: Metal Physics **15**, 1337 (1985).
[34] B. Alling, T. Marten, and I. A. Abrikosov, Phys. Rev. B **82**, 184430 (2010).
[35] D. Gambino and B. Alling, Phys. Rev. B **98**, 064105 (2018).
[36] O. Hegde, M. Grabowski, X. Zhang, O. Waseda, T. Hickel, C. Freysoldt, and J. Neugebauer, Phys. Rev. B **102**, 144101 (2020).
[37] L. Tsetseris, N. Kalfagiannis, S. Logothetidis, and S. T. Pantelides, Phys. Rev. B **76**, 224107 (2007).
[38] E. Mozafari, B. Alling, P. Steneteg, and I. A. Abrikosov, Phys. Rev. B **91**, 094101 (2015).
[39] Z. Zhang, H. Li, R. Daniel, C. Mitterer, and G. Dehm, Phys. Rev. B **87**, 014104 (2013).
[40] G. Abadias, C.-H. Li, L. Belliard, Q. M. Hu, N. Greneche, and P. Djemia, Acta Mater. **184**, 254 (2020).
[41] A. Herwadkar and W. R. L. Lambrecht, Phys. Rev. B **79**, 035125 (2009).

[42] P. Subramanya Herle, M. S. Hegde, N. Y. Vasathacharya, S. Philip, M. V. Rama Rao, and T. Sripathi, J. Solid State Chem. **134**, 120 (1997).
[43] S. Kerdsongpanya, B. Sun, F. Eriksson, J. Jensen, J. Lu, Y. K. Koh, N. V. Nong, B. Balke, B. Alling, and P. Eklund, J. Appl. Phys. **120**, 215103 (2016).
[44] C. X. Quintela, B. Rodríguez-González, and F. Rivadulla, Appl. Phys. Lett. **104**, 022103 (2014).
[45] M. A. Gharavi, S. Kerdsongpanya, S. Schmidt, F. Eriksson, N. V. Nong, J. Lu, B. Balke, D. Fournier, L. Belliard, A. l. Febvrier, C. Pallier, and P. Eklund, J. Phys. D: Appl. Phys. **51**, 355302 (2018).
[46] M. E. McGahay, S. V. Khare, and D. Gall, Phys. Rev. B **102**, 235102 (2020).
[47] A. l. Febvrier, N. V. Nong, G. Abadias, and P. Eklund, Applied Physics Express **11**, 051003 (2018).
[48] N. A. Muhammed Sabeer and P. P. Pradyumnan, Materials Science and Engineering: B **273**, 115428 (2021).
[49] G. Abadias, L. E. Koutsokeras, S. N. Dub, G. N. Tolmachova, A. Debelle, T. Sauvage, and P. Villechaise, J. Vac. Sci. Technol. A **28**, 541 (2010).
[50] A. l. Febvrier, J. Jensen, and P. Eklund, J. Vac. Sci. Technol. A **35**, 021407 (2017).
[51] B. Bakhit, D. Primetzhofer, E. Pitthan, M. A. Sortica, E. Ntemou, J. Rosen, L. Hultman, I. Petrov, and G. Greczynski, J. Vac. Sci. Technol. A **39**, 063408 (2021).
[52] G. Kresse and J. Furthmüller, Phys. Rev. B **54**, 11169 (1996).
[53] G. Kresse and J. Furthmüller, Computational Materials Science **6**, 15 (1996).
[54] P. E. Blöchl, Phys. Rev. B **50**, 17953 (1994).
[55] G. Kresse and D. Joubert, Phys. Rev. B **59**, 1758 (1999).
[56] S. L. Dudarev, G. A. Botton, S. Y. Savrasov, C. J. Humphreys, and A. P. Sutton, Phys. Rev. B **57**, 1505 (1998).
[57] K. Balasubramanian, S. V. Khare, and D. Gall, Acta Mater. **159**, 77 (2018).
[58] P. A. Stampe, M. Bullock, W. P. Tucker, and J. K. Robin, J. Phys. D: Appl. Phys. **32**, 1778 (1999).
[59] Z. Hui, X. Tang, R. Wei, L. Hu, J. Yang, H. Luo, J. Dai, W. Song, X. Liu, X. Zhu, and Y. Sun, RSC Advances **4**, 12568 (2014).
[60] D. Gall, C.-S. Shin, T. Spila, M. Odén, M. J. H. Senna, J. E. Greene, and I. Petrov, J. Appl. Phys. **91**, 3589 (2002).
[61] M. A. Gharavi, G. Greczynski, F. Eriksson, J. Lu, B. Balke, D. Fournier, A. le Febvrier, C. Pallier, and P. Eklund, J. Mater. Sci. **54**, 1434 (2019).
[62] M. Magnuson, M. Mattesini, N. V. Nong, P. Eklund, and L. Hultman, Phys. Rev. B **85**, 195134 (2012).
[63] L. Chaput, P. Pécheur, J. Tobola, and H. Scherrer, Phys. Rev. B **72**, 085126 (2005).
[64] H. I. Yoo, M. W. Barsoum, and T. El-Raghy, Nature **407**, 581 (2000).